\documentclass[twocolumn,prl,amsmath,amssymb,showpacs,superscriptaddress,floatfix]{revtex4}
\usepackage{graphicx}
\usepackage{bm}
\usepackage{lineno,hyperref}
\usepackage{epsf}
\usepackage{xcolor}

\begin{document}
\title{Melting line of silicon modelled with a machine-learning potential}

\author{Yu. D. Fomin \footnote{Corresponding author: fomin314@mail.ru}}
\affiliation{Vereshchagin Institute of High Pressure Physics,
Russian Academy of Sciences, Kaluzhskoe shosse, 14, Troitsk,
Moscow, 108840, Russia }

\begin{abstract}
In the present study we investigate the phase diagram of silicon within the framework of SNAP machine learning
potential model. We show that the melting line of diamond phase of silicon is a linear function of pressure,
which is in good agreement with experimental data. At the same time the melting temperature is strongly underestimated.
Also, this model fails to predict the high pressure phases of silicon.
\end{abstract}

\date{\today}


\maketitle


\section{Introduction}

Similar to carbon, silicon demonstrates complex behavior. The phase diagram of silicon contains
several distinct crystalline phases. At ambient pressure it crystallizes into diamond structure which
transforms into $\beta$-tin at elevated pressures \cite{kubo,voronin}. The melting line of diamond phase of silicon has
negative slope, i.e., $\frac{\partial T}{\partial P}<0$ along the melting line \cite{kubo}. Such behavior often
leads to complex behavior in liquid phase including numerous anomalies, such as density anomaly
(negative thermal expansion coefficient $\alpha_P=\left( \frac{1}{V} \frac{\partial V}{\partial T} \right)_P$),
diffusion anomaly (the diffusion coefficient increases under isothermal compression) and structural
anomaly (the liquid becomes less structures under isothermal compression) \cite{ufn-cs}. Such anomalies were
reported in liquid silicon within the framework of molecular simulations methods with
well recognized Stillinger-Weber (SW) potential \cite{sastry}.

The SW potential is an empirical potential specially developed to model silicon \cite{sw}. Later on it was parameterised
for other substances, such as carbon \cite{sw-c}, germanium \cite{sw-ge} and even water \cite{sw-wat}. Other parameterizations of the SW potential
for silicon are also available in the literature, which better reproduce the properties of silicon
in some special cases. For instance, in Ref. \cite{sw-amorph} a parametrization of SW potential for amorphous silicon
is given.

As it was mentioned above, diamond phase of silicon demonstrates negative slope of the melting line. This is an
important and unusual property which is well reproduced by the SW potential \cite{dozhdikov}. Moreover, the SW potential
with original set of parameters reproduces the melting line of the diamond phase in close agreement with
experimental data. At the same time, the electronic structure of liquid silicon should be different
from the one of the crystalline one, which make the application of a single empirical model to both
crystalline and liquid phases suspicious.

Nowadays more powerful models to describe the interaction potential of a system are available: the ones
based on the machine learning potentials \cite{ml}. Several publicly available potentials of silicon can be found
in special databases. One of the most common machine-learning potentials is based on so-called GAP model.
The phase diagram of silicon obtained with GAP potential resembles some principle features of the experimental
phase diagram: negative slope of the melting line of diamond phase, $\beta$-Sn phase at elevated pressures
and the presence of sh phase \cite{gap-silicon}. At the same time, the agreement of the phase diagram from GAP model and the experimental
one is not satisfactory. The melting temperature of GAP silicon as a function of pressure is systematically lower than the experimental one.
The boundaries of $\beta$-Sn and sh phases are also far from the experimental ones (see Fig. 10 of Ref. \cite{gap-silicon}).

In the present letter we check whether SNAP machine learning potential is able to reproduce the experimental phase diagram of silicon.

\section{System and Methods}

The present work consists of two parts. In the first part we calculate the melting line of silicon with SNAP machine-learning
interaction potential from Ref. \cite{snap-silicon}. The melting line was calculated by two phase method. A rectangular box with 10 lattice units
in the a and b axis and 20 units in the c one was constructed. The lattice constant of the lattice is $a=5.43$ $\AA$, which
corresponds to the lattice constant of silicon at ambient pressure. The lower half of the box was considered as a solid part. It was
equilibrated at $T=500$ K. The upper part was a liquid one and it was melted at $T=5000$ K. After that both parts were simulated for
500 ps with the time step $dt=0.0005$ ps. The system was simulated at constant temperature and constant pressure. A set of pressures was studied:
$P=1$ bar, $10$, $20$, ..., $90$ kbar. Originally the temperatures from $T=800$ K up to $T=1500$ K were studied to roughly localize the
melting point. When the melting point was roughly estimated more simulations with the step in temperature $dT=20$ K were performed. It
allowed us to calculate the melting temperature at given pressure with the accuracy of 20 K.

In the second part of the work we calculate the stable crystal structure of silicon with SNAP potential. We employ Calypso package in
combination with lammps. 20 evolutionary steps with 30 structures in each generation were performed. A list of 50 structures with the
best energies was obtained at each pressure, which allowed to see the ground state and a set of possible metastable structures.

\section{Results and discussion}

\subsection{Melting line}

\begin{figure}

\includegraphics[width=8cm, height=6cm]{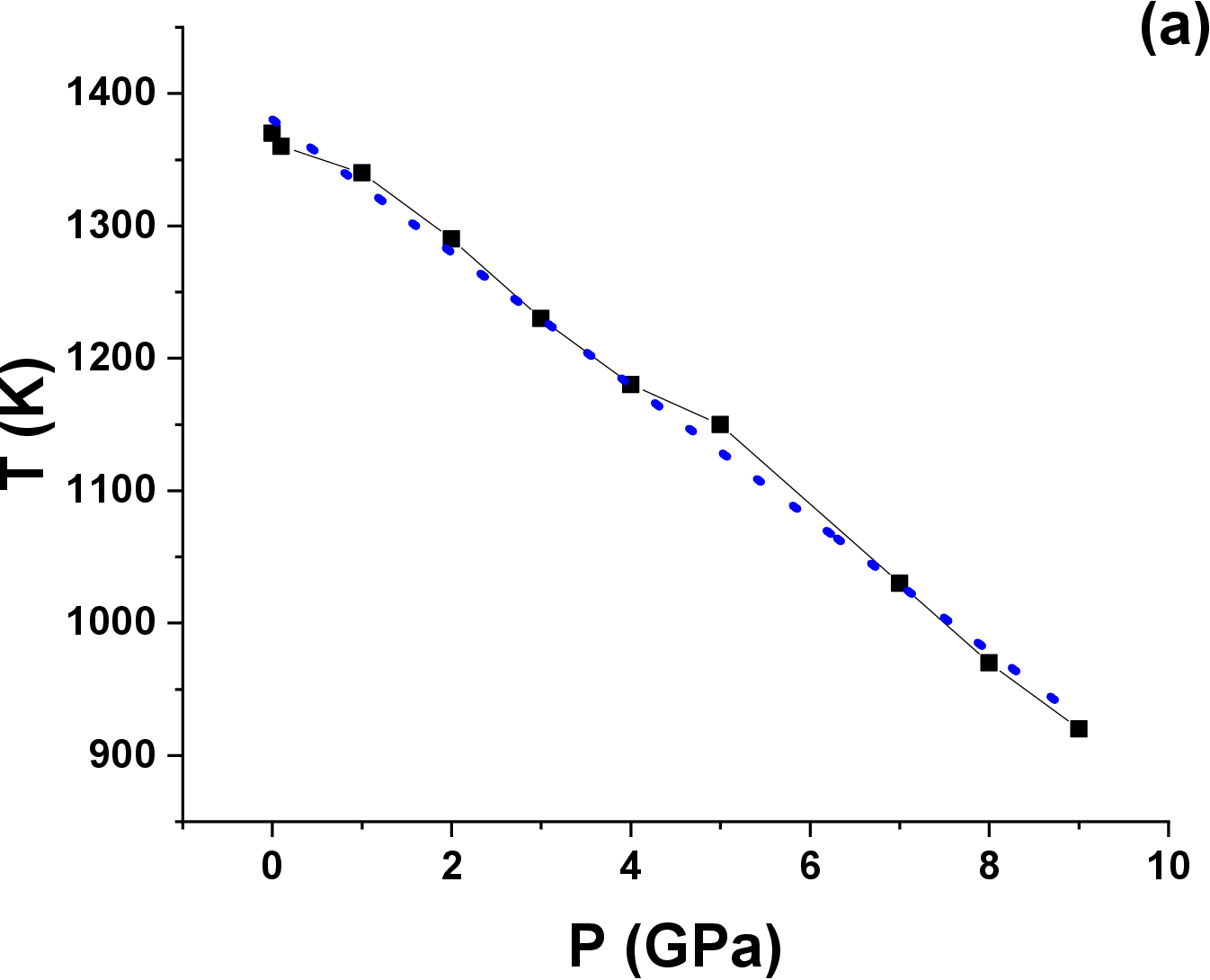}%

\includegraphics[width=8cm, height=6cm]{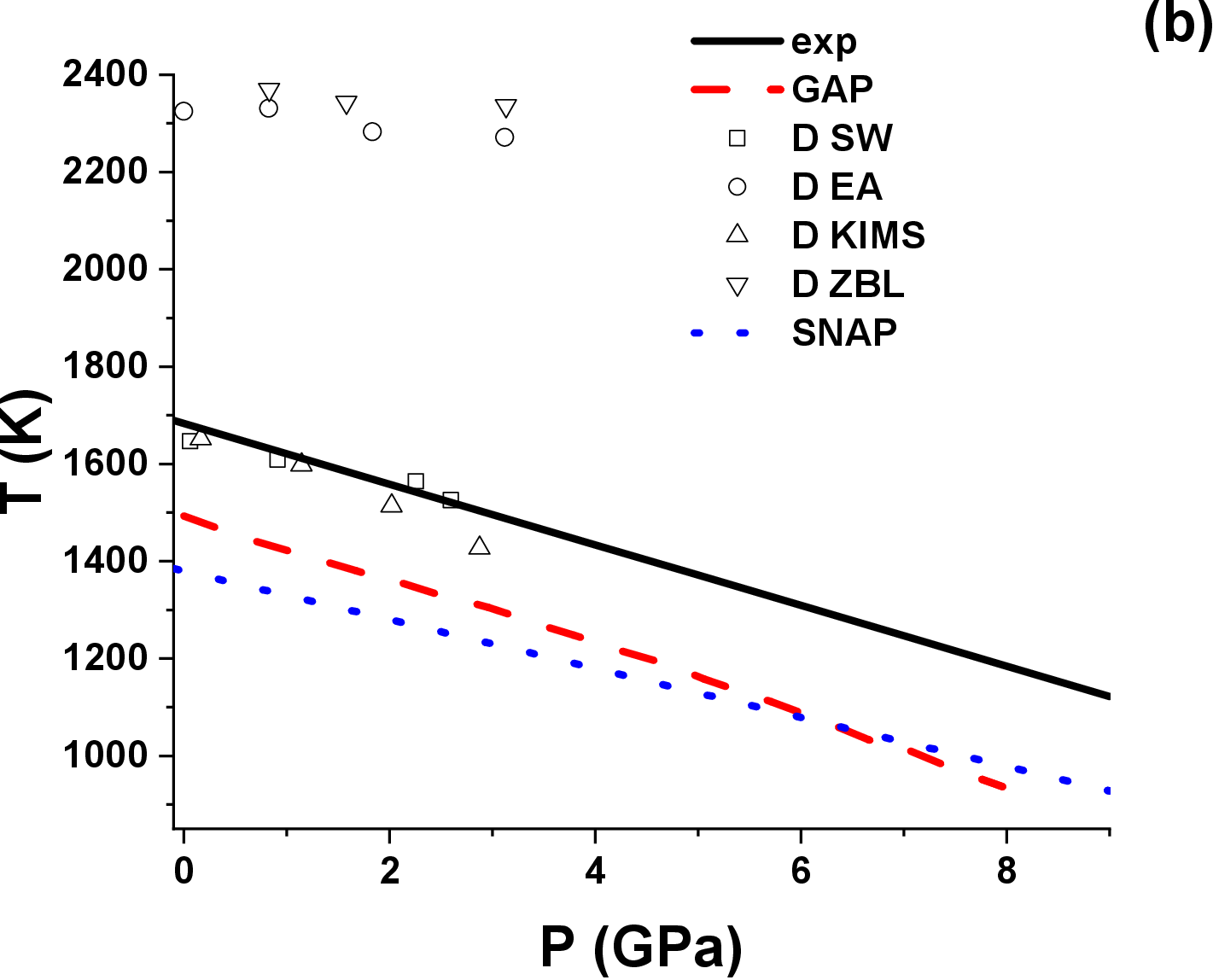}%

\caption{(a) Melting line of SNAP model of silicon. The symbols show raw data. The line is the linear fit of these data. (b) A comparison of the data for SNAP model
with the literature ones. The curve 'exp' means the experimental curve from Ref. \cite{kubo}. The symbols 'D SW', 'D EA', 'D KIMS' and 'D ZBL' refer
to the results for SW, EA, KIMS and ZBL models reported by Dozhdikov and coauthors in Ref. \cite{dozhdikov}. In the case of SNAP model we give only the linear fit of the data.}
\label{ml}
\end{figure}

In Fig. \ref{ml} (a) we show the data for the melting line of SNAP model of silicon. The data are well fitted to a line
$T(K)=-50.23P(GPa)+1380.39(K)$. This result is compared with the literature data in Fig. \ref{ml} (b).
Experimental data on the phase diagram of silicon in a wide range of pressures were reported in Refs. \cite{kubo,voronin}.
In particular, as it was shown in Ref. \cite{kubo} experimental melting line of diamond phase of silicon can be fitted
by a straight line $T(K)=-62.3P(GPa)+1683(K)$. It means that (i) SNAP model does reproduce the negative slope of the melting
line and linear dependence of the melting temperature on pressure, but (ii) SNAP model strongly underestimates the melting
temperature and the slope $\frac{dT}{dP}$.

Figure \ref{ml} (b) shows also a comparison with the results for the GAP model \cite{gap-silicon} and four empirical models: SW, EA, KIHS
and ZBL. All data for
the empirical models are taken from Ref. \cite{dozhdikov}. It is seen that the GAP model also underestimates the melting point.
A comparison of the GAP model with DFT calculations is given in the original paper \cite{gap-silicon}. It is shown
that within the framework of employed functionals DFT also underestimates the melting point at ambient pressure (see Fig. 10
of Ref. \cite{gap-silicon}). At the same time the melting line of the GAP model is closer than the one of the SNAP one.

The empirical SW model demonstrates the best agreement with experiment. The results for the KIHS model also look reasonable at low pressures,
but deviate to lower temperatures at the elevated ones. Two other models (EA and ZBL) strongly overestimate the melting temperatures at all pressures.

One can make several conclusions from the results of Fig. \ref{ml} (b). First, both machine-learning models underestimate the melting point. This is
related to the underestimation of the melting point of silicon in DFT calculations (see Fig. 10 of Ref. \cite{gap-silicon}). Silicon is a text-book
example of the DFT method with well developed methodology for calculation of different properties. Surprisingly, this methodology has not
been implemented to calculate the melting line of silicon with reasonable accuracy. Such implementation and construction of an
accurate machine-learning potential for liquid silicon would be of strong interest.

Another conclusion is that currently available machine-learning potentials for silicon do not give better accuracy comparing to the empirical models.
At the same time a comparison of different experimental models shows that they demonstrate strongly different results, which means that
these models are not trustable.

\subsection{Crystal structure calculation}

In the second part of the work we calculated the ground state crystal structure of SNAP model of silicon. At the ambient pressure
the stable crystal structure of the system is diamond with the ground state energy $E_g=-5.41838$ eV per atom. The diamond structure
appears to be the ground state up to the pressure $40$ GPa. At the same time, at pressure about $12$ GPa the relative difference between
the energy of the diamond structure and the $\beta$-Sn one becomes extremely small: of the order of $0.05 \%$. Such small differences
in energy cannot be resolved either in molecular dynamics simulation, or in the lammps minimization process. For this reason, one
can say, that at pressures about $12$ GPa a trnsformation of the diamond phase into the $\beta$-Sn one is possible. However, strictly speaking,
the energy of the diamond phase is always a bit lower than the $\beta$-Sn one.

\begin{figure}

\includegraphics[width=8cm, height=6cm]{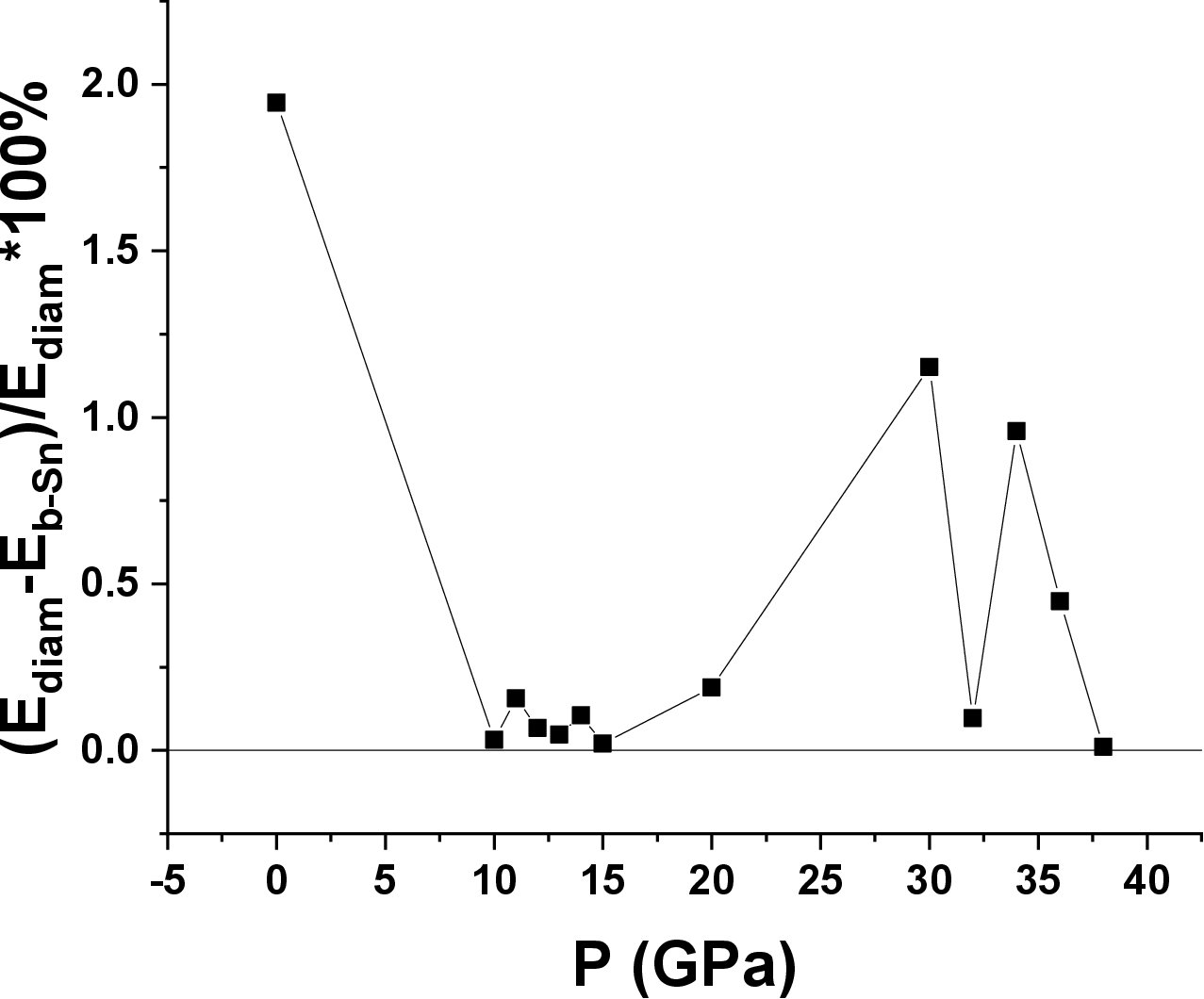}%

\caption{Relative difference between the energy of the diamond structure and $\beta$-Sn one for the SNAPP model of silicon as a function of pressure.}
\label{eps}
\end{figure}

According to the experimental phase diagram \cite{kubo}, when pressure is further elevated, silicon transforms into orthorombic (pressure
about $15$ GPa) and then to a simple hexagonal phase (sh) phase ($P \approx 16.5$ GPa). Both these phases are not obtained
in the ground state search with SNAP model. The next stable phase is $P6_3/mmc$ (group 194) and it appears at $P=40$ GPa.

From the discussion above one can conclude that SNAP model fails to describe the high pressure phases of carbon. For this reason, we do not
calculate the phase transformation lines of these phases.

\section{Conclusions}

Liquid silicon is an interesting liquid with numerous anomalous properties \cite{sastry}. However, as it was stated above,
the results of different empirical models can strongly deviate from each other. At the same time, silicon can be efficiently studied
in DFT calculations. There are several interaction potential models of silicon based on machine learning methods. These
models correctly predict that the melting line of diamond phase of silicon is a straight line in a wide range of pressures,
but they strongly underestimate the melting temperature. For this reason, we expect that the results of these models
for liquid silicon are also questionable and, in the best case, give only qualitatively correct results. The overall
performance of GAP model, as it was reported in Ref. \cite{gap-silicon} is better than the SNAP one considered in the
present study.

Taking into account strong interest to liquid silicon and well established DFT methodology for this element
we believe that development of an accurate machine-learning potential for liquid silicon should be performed in a near future.

\bigskip

This work was carried out using computing resources of the federal
collective usage center "Complex for simulation and data
processing for mega-science facilities" at NRC "Kurchatov
Institute", http://ckp.nrcki.ru, and supercomputers at Joint
Supercomputer Center of the Russian Academy of Sciences (JSCC
RAS). 

\end{document}